\documentclass[a4, 12pt]{article}
\usepackage[margin=0.5in]{geometry}
\usepackage{graphicx}

\title{Variational autoencoders for tissue heterogeneity exploration from (almost) no preprocessed mass spectrometry imaging data.}
\author{Paolo Inglese, James L. Alexander, Anna Mroz, Zoltan Takats, Robert Glen}

\begin{document}

\maketitle

\begin{abstract}
The paper presents the application of Variational Autoencoders (VAE) for data dimensionality reduction and explorative analysis of mass spectrometry imaging data (MSI). The results confirm that VAEs are capable of detecting the patterns associated with the different tissue sub-types with better performance than standard approaches.
\end{abstract}


\section{Introduction}

In recent years, mass spectrometry imaging has acquired increasing importance among the metabolic analytical techniques, proving to be able to capture and represent in a concise way the complex heterogeneous biochemical interactions.
Because of its native high dimensionality (often of the order of $10^3-10^4$) and the large number of samples (pixels), technical difficulties arise in the exploratory analysis of untargeted datasets, such as cancer specimens, for which biochemical patterns are still unknown. In this context, data dimensionality reduction can represent an opportune strategy with the aim of detecting and summarising the underlying patterns that characterise the data structure.

Among the different available techniques, linear methods such as PCA or MDS, still widely used in the field of mass spectrometry imaging, have shown to be limited in capturing complex non-linear molecular patterns that can characterise the highly heterogeneous tissues \cite{thomas2016dimensionality}. For those reasons, we apply deep generative VAE models to perform dimensionality reduction, data visualisation, and data compression, proving that they perform better than standard methods, such as PCA. VAE deep structure models can capture the distributions of the latent variables that determine the observed patterns and, at the same time, provide a powerful tool to reconstruct the original data and to generate unseen data that has resemblance of the original data.

Specifically, under the assumption that different tissue types are characterised by significantly distinct mass spectra profile patterns (even if unmatched), we show that VAEs are capable of retrieving the latent space representation in a reasonable amount of time with no need for either peak detection or spectral alignment.

\section{Variational autoencoders}

Here we report a brief summary of the Variational Autoencoder (VAE) algorithm\footnote{For a detailed description of the algorithm, we refer the reader to the original paper \cite{2013arXiv1312.6114K}.}.

VAE  \cite{2013arXiv1312.6114K} represents an extension of Bayesian variational inference to large scale datasets where standard approaches (such as MCMC sampling) are infeasible. The purpose of VAE is to learn an approximate posterior inference model with continuous latent variables.
 
Let $\mathbf{X}=\left\{\mathbf{x}^{(i)}\right\}_{i=1,\ldots,N}$ be a dataset of $N$ i.i.d. samples of a continuous random variable $\mathbf{x}$. The assumption is that the data samples are generated by a random process, involving a "latent" random variable $z$. Specifically, $\mathbf{z}$ is sampled from a prior distribution $p_{\mathbf{\theta}^*}(\mathbf{z})$, and $\mathbf{x}$ is sampled from the conditional distribution $p_{\theta^*}(\mathbf{x}|\mathbf{z})$. We must stress that the latent variable $\mathbf{z}$ and the parameter $\mathbf{\theta}^*$ are hidden from the direct observation.

The estimation of the parameter $\theta$ and the posterior inference of the latent variable $z$ is performed by introducing a recognition model $q_{\phi}(\mathbf{z}|\mathbf{x})$ which acts as an \emph{encoder} that tries to approximate the true posterior $p_{\theta}(\mathbf{z}|\mathbf{x})$. Analogously, after sampling a $\mathbf{z}$ from the recognition model, a probabilistic \emph{decoder} can reconstruct the possible corresponding sample $\mathbf{x}$.

On of the key aspects of the VAE algorithm is the "reparameterisation trick", by which we restrict the recognition model to be differentiable. In this way, learning schemes such as gradient descent can be applied. In our model, following the original method, $\mathbf{z}$ is sampled from a normal distribution with diagonal variance and defined as $\mathbf{z=\mu + \sigma\epsilon}$, where $\mathbf{\mu}$ and $\mathbf{\sigma}$ are the learnt mean and standard deviations, respectively, and $\mathbf{\epsilon} \sim \mathcal{N}(0, I)$.

For the VAE applied in our work, the prior $p_{\mathbf{\theta}}(\mathbf{z}) = \mathcal{N}(\mathbf{z}; 0, I)$ and the posterior of the latent variable $q_{\phi}(\mathbf{z}|\mathbf{x}) = \mathcal{N}(\mathbf{z}; \mathbf{\mu}, \mathbf{\sigma}^2 \mathbf{I})$ were both assumed to be approximately a Gaussian with a diagonal covariance (mean $\mathbf{\mu}$ and variance $\mathbf{\sigma}^2$ are outputs of a multi-layer perceptron (MLP) applied to the data points $\mathbf{x}^{(i)}$), and the posterior inference model $p_{\theta}(\mathbf{x}|\mathbf{z})$ was assumed to be a multivariate Bernoulli distribution applied to the $\mathbf{z = \mu + \sigma \epsilon}$, with $\mathbf{\epsilon} \sim \mathcal{N}(0, \mathbf{I})$. In this case, the cost function $\mathcal{L}$ is defined as the sum of the Kullback-Leibler divergence from $p_\theta(\mathbf{z})$ to $q_\phi(\mathbf{z}|\mathbf{x})$ and the negative reconstruction error\footnote{$J$ is the latent space dimensionality.},
\begin{equation}
	\mathcal{L}(\mathbf{\theta, \phi, x}^{(i)}) = \frac{1}{2}\sum_{j=1}^J \left( 1 + \log \left( (\sigma_j^{(i)})^2\right)  - (\mu_j^{(i)})^2 - (\sigma_j^{(i)})^2\right) + \frac{1}{L} \sum_{l=1}^L \log p_{\theta} (x^{(i)}|z^{(i,l)})
\end{equation}
where $\mathbf{z}^{(i,l)} = \mathbf{\mu}^{(i)} + \mathbf{\sigma}^{(i)} \odot \mathbf{\epsilon}^{(l)}$ and $\mathbf{\epsilon} \sim \mathcal{N}(0, \mathbf{I})$.

\begin{figure}
	\centering
	\includegraphics[width=0.5\textwidth]{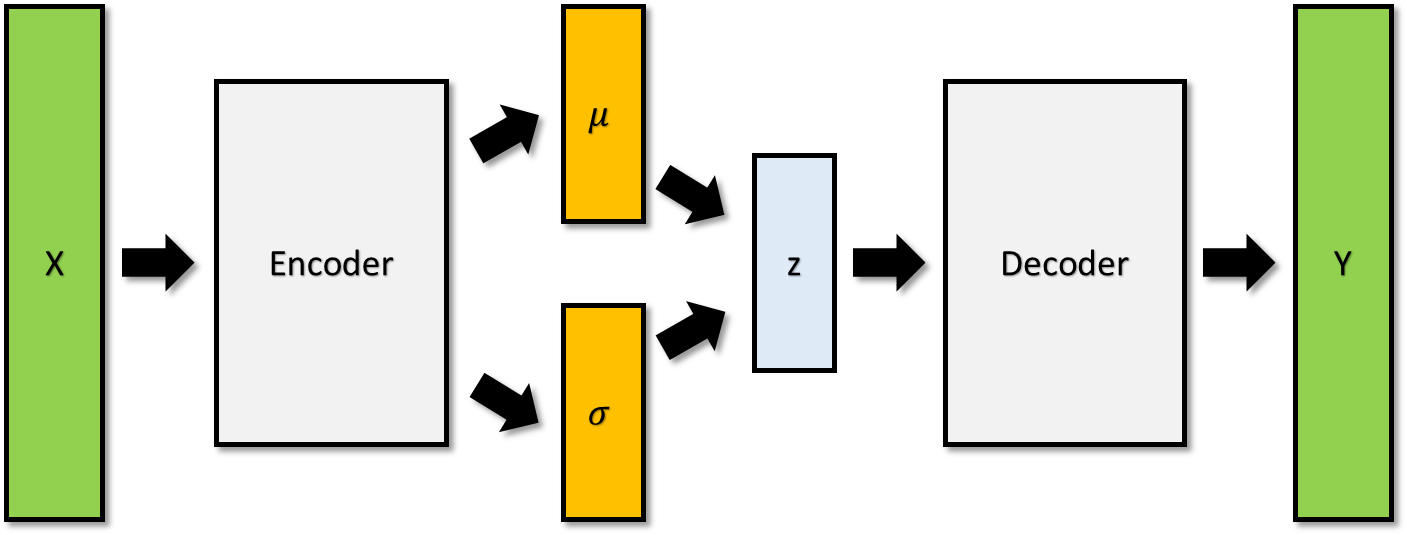}
	\caption{\footnotesize{"A scheme of a variational autoencoder. The encoder MLP transforms the input into the mean and the log-variance of the prior $p_\theta(\mathbf{z})$. Using the reparameterization trick, $\mathbf{z}=\mathbf{\mu} + \mathbf{\sigma} \odot \epsilon$ is therefore passed to a decoder MLP which finally generates the reconstructed data sample $\mathbf{y}$.}}
	\label{Fig:vae}
\end{figure}

For Bernoulli outputs, the second term corresponds to the cross-entropy
\begin{equation}
	\log p(x|z) = \sum_{i=1}^M x_i \log y_i + (1 - x_i) \log(1 - y_i)
\end{equation}
where $\mathbf{y}$ is the output of an MLP.
A schematic representation of the network architecture is shown in Fig. \ref{Fig:vae}.

\section{Experiments}

\begin{figure}[h!]
	\centering
	\includegraphics[width=0.8\textwidth]{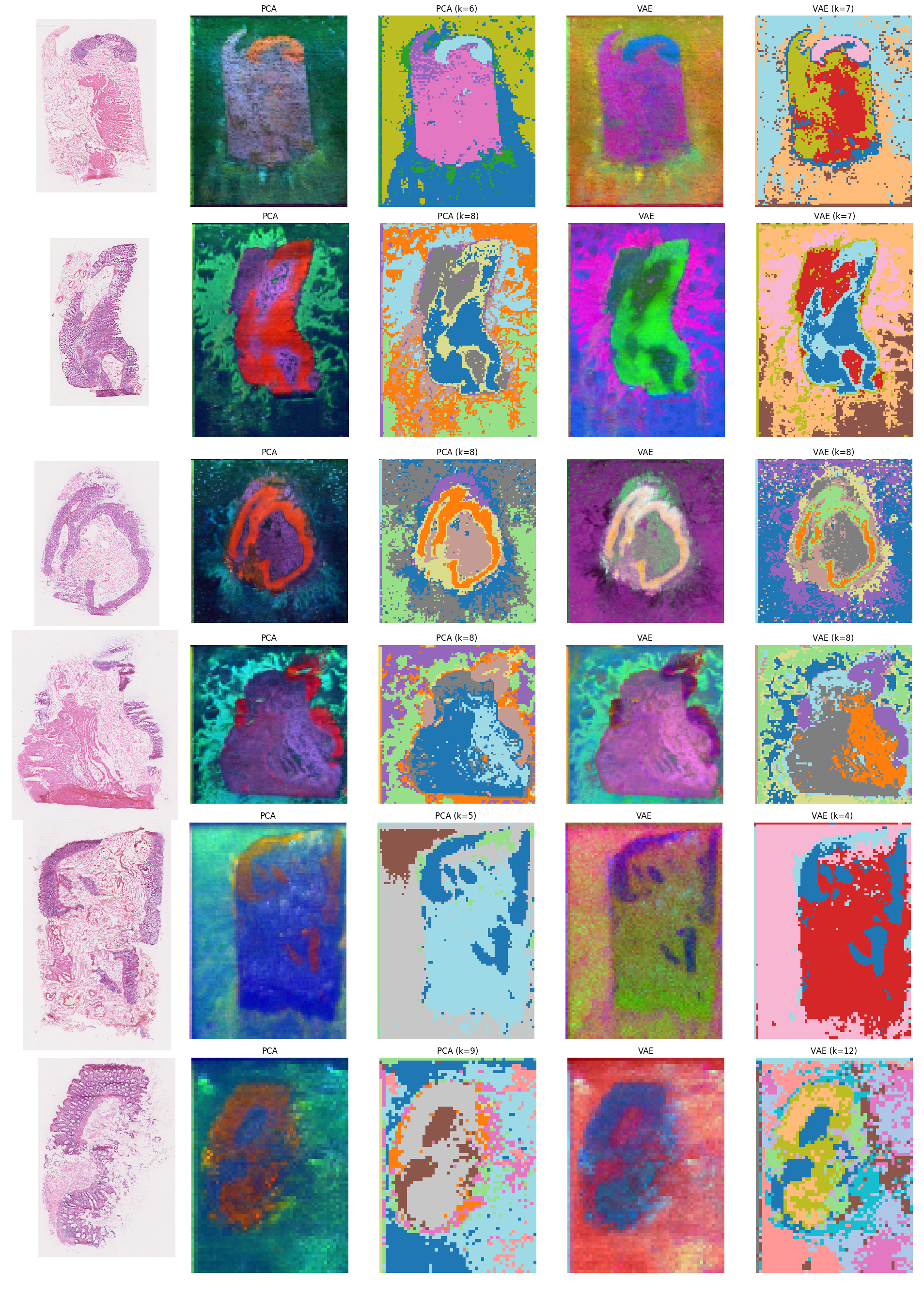}
	\caption{\footnotesize{In each row, the results of one of the tested samples are reported. From left to right: H\&E stained tissue samples, RGB visualisation of the normalised scores of the 3 first principal components, GMM clustering using the PCA scores, RGB visualisation of the normalised 3 latent variables of the VAE model, GMM clustering using the VAE latent variables. The selected number of clusters corresponded to the closest partition to the optical image (in a range of 2 to 19).}}
	\label{Fig:all}
\end{figure}

Desorption electrospray ionisation mass spectrometry (DESI) imaging data from 6 human colon tissue specimens (first column of Fig. \ref{Fig:all}) was used to evaluate the capabilities of VAE models to perform data compression (dimensionality reduction + data reconstruction), compared to a standard PCA. All the mass spectrometry data were acquired using a Waters  XEVO-G2XS Q-TOF mass spectrometer, set in the negative ion mode.

All the spectral profiles were linearly interpolated with a uniform sampling step of 0.02 m/z in the range of 600-1,000 m/z, resulting in a dimensionality of 20,001. The only preprocessing steps necessary for our analysis consisted of: 1) application of the $\log(\mathbf{x} + 1)$ intensity transformation to reduce data heteroscedasticity; 2) intensities transformation by the logistic function on the median centred spectrum intensities

\begin{equation}
\mathbf{x'}^{(i)} = \frac{1}{1 + \exp (-(\mathbf{x}^{(i)} - \mathbf{m}^{(i)}))}.
\end{equation}

where $\mathbf{m}^{(i)}$ is the (non-zero) median of the spectrum intensities.
Those transformations allowed mapping of the spectral intensities onto a range of [0-1], in such a way they could be interpret as probabilities to find a source of signal at the corresponding m/z value, and, at the same time and to compress the large variabilities of high/low intense signals across the different pixels.

The purpose of these experiments is to compare the capability of VAE and PCA to reduce the data dimensionality while preserving the spatial structures in the tissues associated with a molecular heterogeneity, through their latent space representations of the mass spectrometry data.

\begin{figure}[h!]
	\centering
	\includegraphics[width=\textwidth]{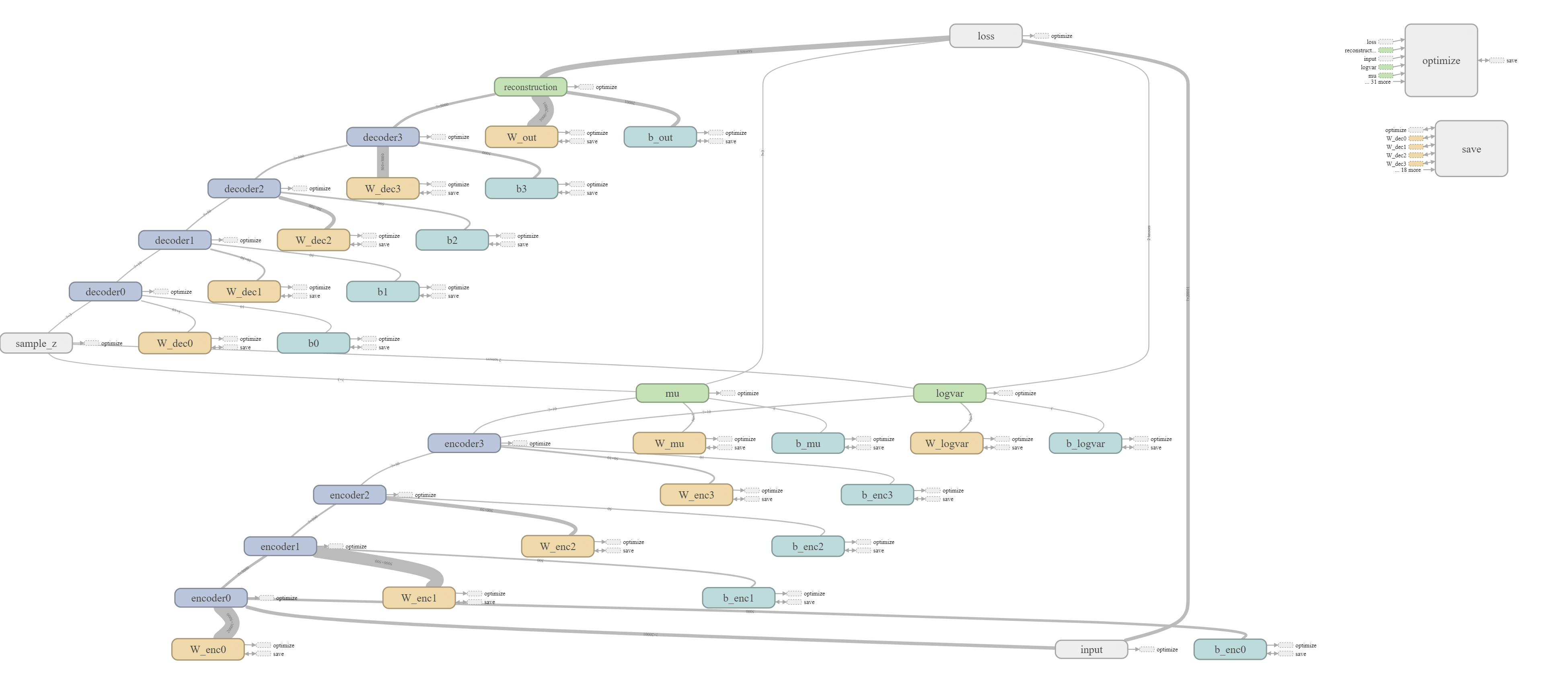}
	\caption{\footnotesize{Schematic representation of the VAE model used in our experiments. The image was generated using Tensorboard.}}
	\label{Fig:model}
\end{figure}

\begin{figure}[h!]
	\centering
	\includegraphics[width=\textwidth]{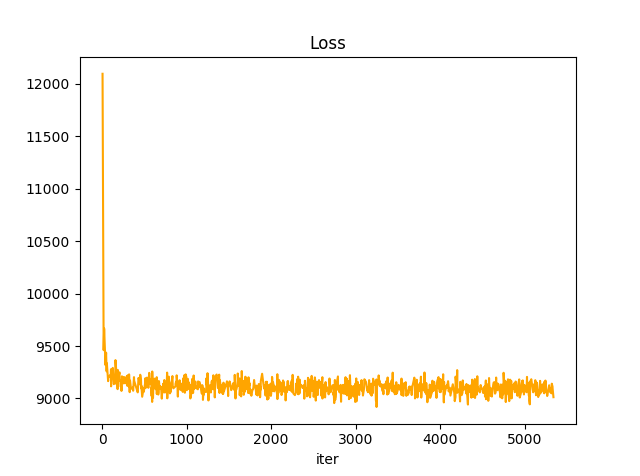}
	\caption{\footnotesize{Loss curve for the training of the VAE model. After few iterations (less than 200) the loss fluctuates around its minimum value.}}
	\label{Fig:loss}
\end{figure}

The used VAE model constructed consisted of 20,001 input neurons, an encoder MLP with 5 hidden layers (5,000, 2000, 500, 50, 10, neurons respectively), a latent dimension of 3 and a decoder MLP with the mirrored topology of the encoder MLP. In each MLP, exponential linear unit (ELU) activations were employed. The network was trained with mini-batches of 256 spectra using the Adam algorithm \cite{2014arXiv1412.6980K} with a learning rate of 0.0001 for 30 epochs (Fig. \ref{Fig:model}). The loss function converged after few iterations (usually less than 200) as shown in Fig. \ref{Fig:loss}.
The VAE model was trained using a leave-one-sample-out procedure. The spectra of 5 tissue samples were used as a training set (corresponding to $\sim$40,000 spectra), and the model was evaluated on the spectra of the held-out tissue sample.
In each round of the evaluation procedure, after training the model, the 3-dimensional latent variables associated to each pixel of the test sample were plotted as (normalised) intensities of the RGB channels. In this way, it could be seen that it was straightforward to visualise how the 3 latent components were mixing across the different regions of the tissue, while expecting that similar tissue types (characterised by similar spectral profiles) would be assigned to similar colours.
As a comparison, the same graphical procedure was used to plot the scores of the first 3 principal components calculated by projecting the test spectra on the loadings obtained from the training set.

All the VAE models were trained and tested on an NVIDIA GTX 1080 Ti using Tensorflow \cite{tensorflow2015-whitepaper} resulting in a computational time of about 6 minutes, whereas PCA was calculated through Scikit-learn package for Python \cite{scikit-learn}.

\begin{figure}[h!]
	\centering
	\includegraphics[width=\textwidth]{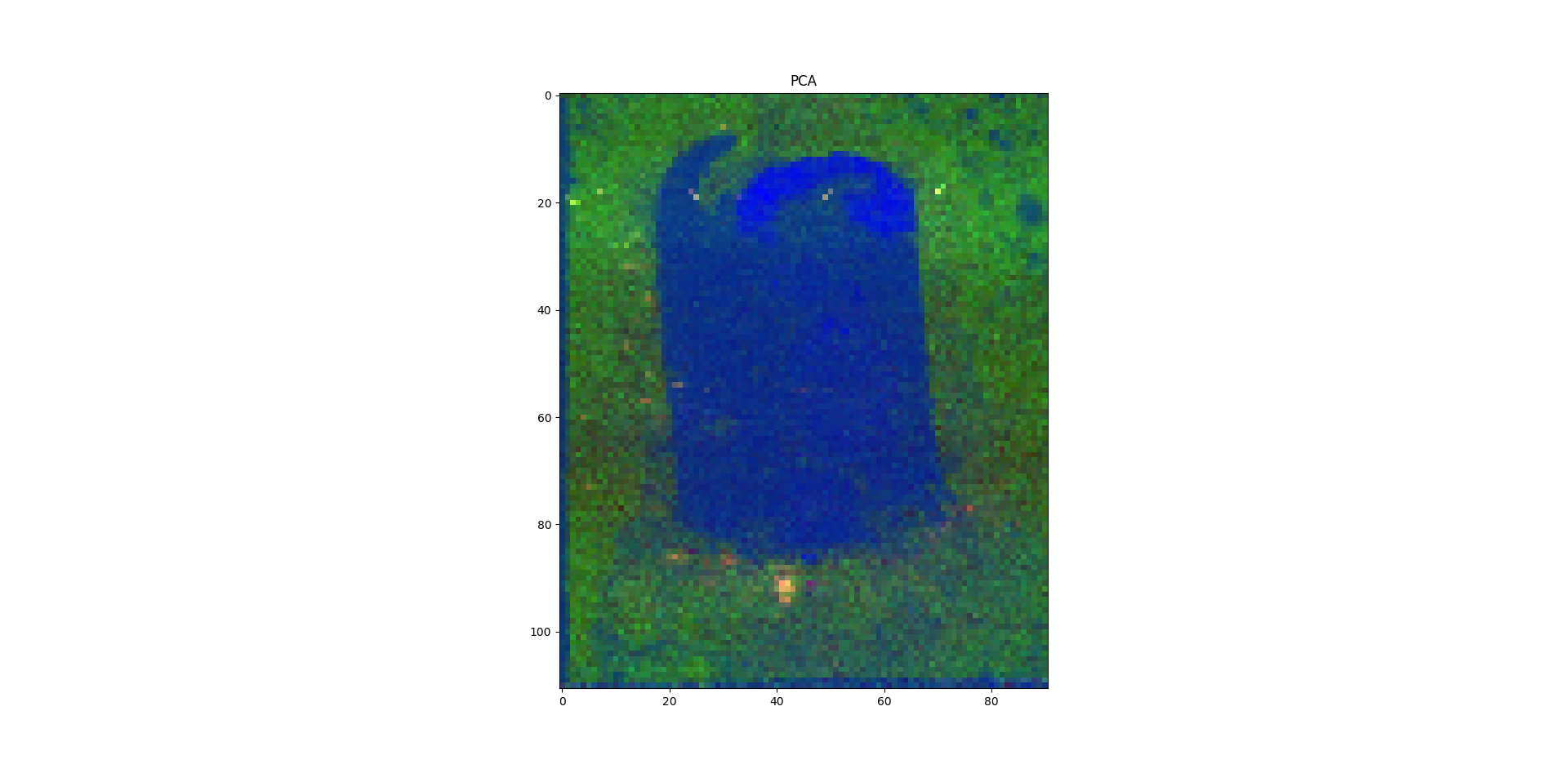}
	\caption{\footnotesize{Visualisation of the first 3 principal components normalised scores of the TIC normalised data, as RGB intensities. Despite the first 3 principal components explained $\sim80\%$ of the total variance, the captured variation consisted mainly of noise, as visible outside of the tissue region.}}
	\label{Fig:pca_var}
\end{figure}

Most of the mass spectrometry signal was uninformative, as shown by the fact that the first 3 principal components explained $~10\%$ of the total variance. Additionally, it was observed that when applying the total-ion-count normalisation followed by the log-transformation, the first 3 principal components could explain up to $80\%$ of the total variance, but the images corresponding to those scores were noisy and the tissue structures were barely distinguishable (Fig. \ref{Fig:pca_var}). For this reason, the same preprocessing described before for both VAE and PCA was used.

\begin{figure}[h!]
	\centering
	\includegraphics[width=\textwidth]{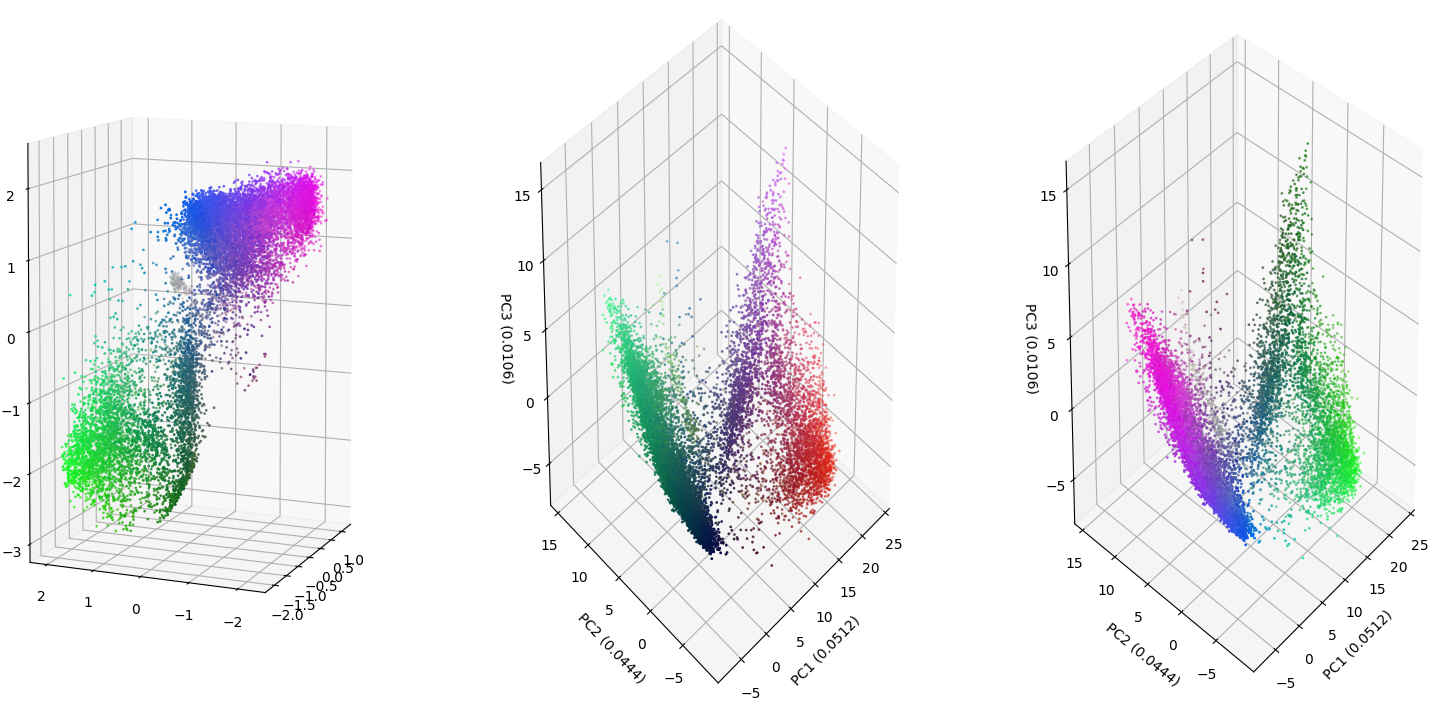}
	\caption{\footnotesize{Left: Scatter plot of the VAE 3-dimensional latent space for the second sample (second row of Fig \ref{Fig:all}) shows the capability of VAE of distinguishing some of the clusters (e.g. the yellow cluster). The point colours correspond to their normalised components (RGB). Center: Scatter plot of the first 3 principal components scores (In each axis, the explained variance is reported). Right: Here the colours of the PCA scores are the same used for the VAE scatter plot, in order to compare the position of same points.}}
	\label{Fig:scatter3d}
\end{figure}

\begin{figure}[h!]
	\centering
	\includegraphics[width=\textwidth]{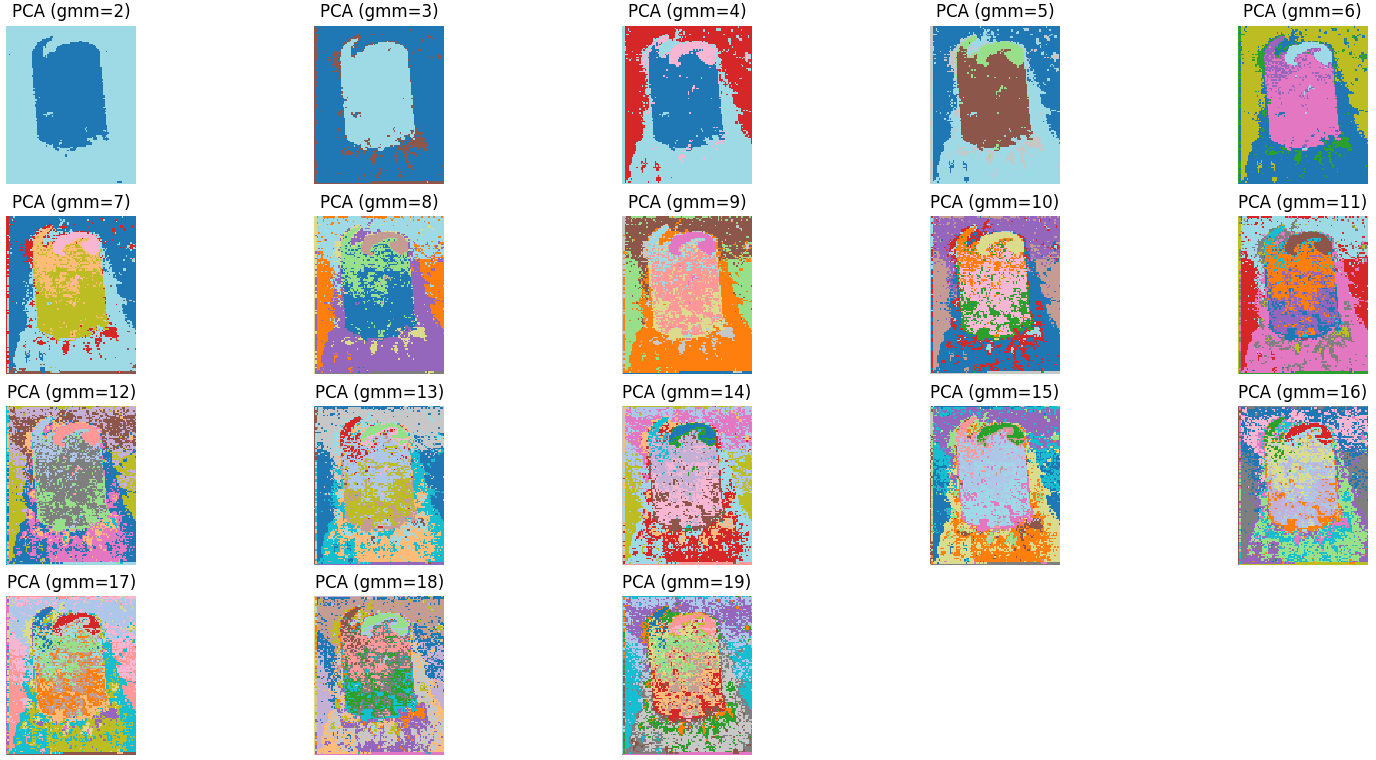}
	\caption{\footnotesize{Example of the GMM clustering with a number of clusters varying in a range of 2 to 19 applied to the first 3 principal components shows that some of the main tissue structures (e.g. the sub-type in the centre, Fig. \ref{Fig:all}) were not detected.}}
	\label{Fig:gmm_pca}
\end{figure}

\begin{figure}[h!]
	\centering
	\includegraphics[width=\textwidth]{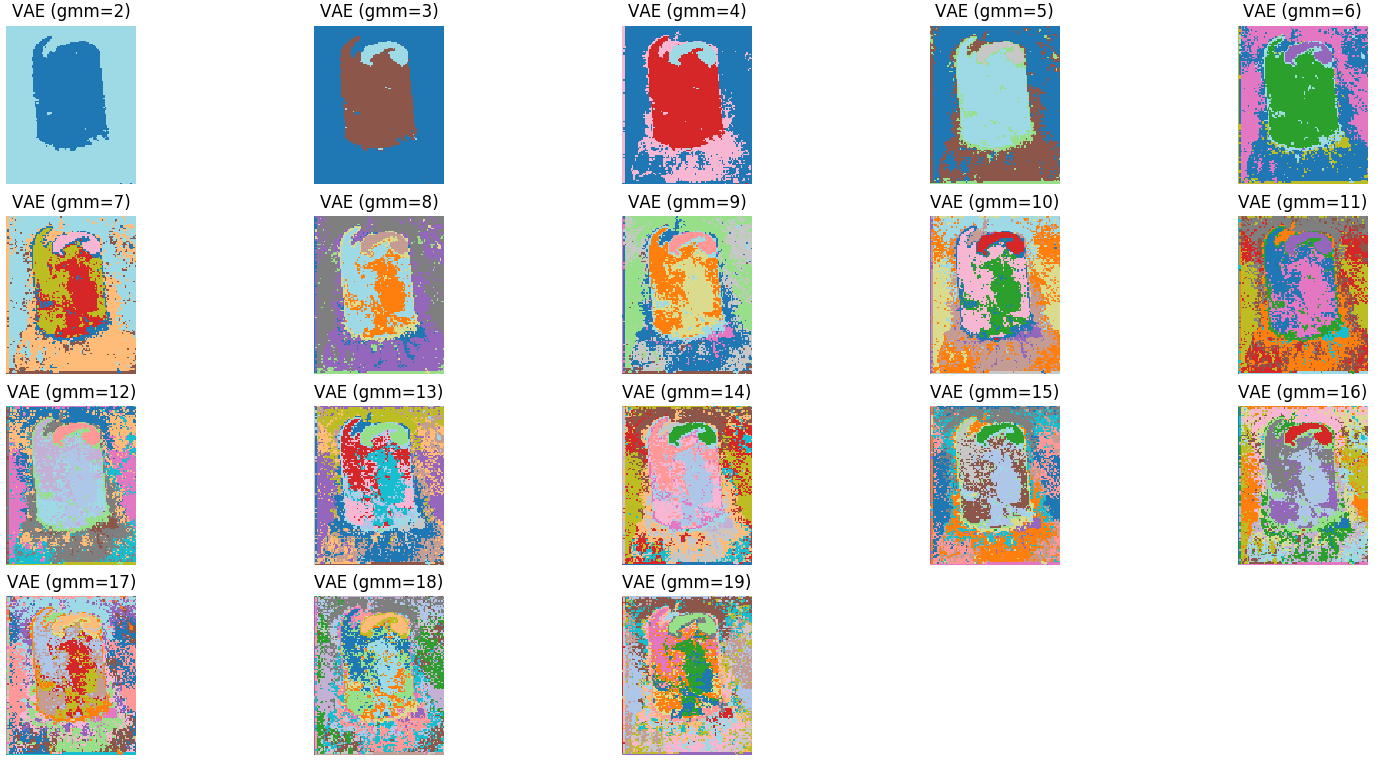}
	\caption{\footnotesize{Example of the GMM clustering with a number of clusters varying in a range of 2 to 19 applied to the VAE latent variables shows that in almost all the partitions, the main tissue structures (e.g. the sub-type in the centre, Fig. \ref{Fig:all}) were identified as different sub-types.}}
	\label{Fig:gmm_vae}
\end{figure}

As reported in Fig. \ref{Fig:scatter3d}, VAEs were capable of differentiating more clusters than PCA which corresponds to distinguishing between a larger number of different substructures (e.g. in the centre of the tissue of the first sample in Fig. \ref{Fig:all}). This result was confirmed by applying a clustering on the latent representation of the test samples using Gaussian Mixture models (GMM) (full covariance, num. repetitions = 5, max. iterations = 1000). Indeed, it was observed that the VAE model could provide clusters that more accurately resembled the tissue morphology than those identified using the corresponding PCA latent representation (Fig. \ref{Fig:all}). An example of the clusters generated by GMM with a varying number of components (2 to 19) using the PCA scores and VAE latent variables is shown in Fig. \ref{Fig:gmm_pca} and Fig. \ref{Fig:gmm_vae}. 

Subsequently, since PCA and VAE can be used as methods to compress the data, both providing a way to reconstruct the original data from their latent (low-dimensional) representation, the accuracy of the reconstructed data evaluated compared to the original input. PCA reconstruction was performed by multiplying the 3-dimensional scores by the corresponding 3 principal loadings of the training set, whereas VAE reconstruction was performed by passing the test sample latent variables through the decoder MLP of the model fitted on the training set. The mean squared error (MSE) (Table \ref{Table:mse}), calculated between the input and reconstructed data, confirmed that the VAE models were capable of compressing and reconstructing the data more faithfully.

\begin{table}
	\centering
	\begin{tabular}{ |c|c|c| } 
		\hline
		Sample & PCA & VAE \\ 
		\hline
		sample1 & 0.06804 & 0.02736 \\ 
		sample2 & 0.06856 & 0.02802 \\ 
		sample3 & 0.06782 & 0.02707 \\
		sample4 & 0.06677 & 0.02735 \\
		sample5 & 0.08209 & 0.04176 \\
		sample6 & 0.06525 & 0.02771 \\
		\hline
	\end{tabular}
	\caption{\footnotesize{Mean squared error between the input and the reconstructed spectral data using PCA and VAE confirms that VAE models can compress and reconstruct the original data more faithfully.}}
	\label{Table:mse}
\end{table}

\section{Conclusions and future work}

Mass spectrometry imaging data represent an invaluable resource of information about the biochemical interactions characterising living organisms. Because of their native high dimensionality, data visualisation and compression are of critical importance in order to understand the underlying patterns representing possible metabolic pathways. Compared to standard methods, such as PCA, deep Variational Autoencoders represent a powerful tool for the unsupervised data exploration of such data, being able to capture the latent variables distribution. Using Variational autoencoders on DESI imaging data, we have shown that this method can compress the data better than PCA, and provide more precise clustering results.

In future we will test the performances of VAE models for the dimensionality reduction of MSI dataset from heterogeneous tissue types and acquired with different platform, in order to evaluate the limit of applicability of this technique.

\section{Acknowledgements}
Human tissue samples were obtained with informed consent under local ethical approval (14/EE/0024).

\bibliographystyle{plain}
\bibliography{paper}

\end{document}